\documentclass[twocolumn,showpacs,nofootinbib]{revtex4}

\include{amssymbol}

\usepackage{color}

\def\AJ{{\it Ap. J.} }

\def\CQG{{\it Class. Quantum Gravity} }

\def\PL{{\it Phys. Lett.} }
\def\PR{{\it Phys. Rev.} }
\def\PRL{{\it Phys. Rev. Lett.} }

\def\ep{\epsilon}   
   \def\ka{\kappa}
   
\def\si{\sigma}   
   
\def\om{\omega}  \def\De{\Delta} 
   
 \def\Om{\Omega} \def\mn{{\mu\nu}}

 \def\frac#1#2{{\textstyle{{#1}\over
{#2}}}} 
\def\lsim{\mathrel{\rlap{\lower4pt\hbox{\hskip1pt$\sim$}}
\raise1pt\hbox{$<$}}}
\def\gsim{\mathrel{\rlap{\lower4pt\hbox{\hskip1pt$\sim$}}
\raise1pt\hbox{$>$}}} \def\sqr#1#2{{\vcenter{\vbox{\hrule height.#2pt
\hbox{\vrule width.#2pt height#1pt \kern#1pt \vrule width.#2pt} \hrule
height.#2pt}}}}
\def\square{\mathchoice\sqr66\sqr66\sqr{2.1}3\sqr{1.5}3}
\def\beq{\begin{equation}} \def\eeq{\end{equation}}
\def\beqa{\begin{eqnarray}} \def\eeqa{\end{eqnarray}}
\def\eq#1{Eq. (\ref{#1})}

\begin{document}

\title{Reheating via a generalized non-minimal coupling of curvature to matter}

\author{Orfeu Bertolami\footnote{Also at Instituto de Plasmas e F\'isica Nuclear, Instituto Superior T\'ecnico, Av. Rovisco Pais, 1, 1049-001, Lisboa Portugal.}}
\email{orfeu.bertolami@fc.up.pt}
\affiliation{Departamento de F\'{\i}sica e Astronomia, Faculdade de Ci\^encias, Universidade do Porto,\\Rua do Campo Alegre 687,
4169-007 Porto, Portugal}

\author{Pedro Fraz\~ao}
\email{pedro.frazao@ist.utl.pt}

\author{Jorge P\'aramos}
\email{paramos@ist.edu}
\affiliation{Instituto de Plasmas e Fus\~ao Nuclear, Instituto Superior T\'ecnico\\Av. Rovisco Pais 1, 1049-001 Lisboa, Portugal}

\date{\today}

\begin{abstract}
In this work one shows that a generalized non-minimal coupling between geometry and matter is compatible with Starobinsky inflation and leads to a successful process of preheating, a reheating scenario based on the production of massive particles via parametric resonance. The model naturally extends the usual preheating mechanism, which resorts to an {\it ad-hoc} scalar curvature-dependent mass term for a scalar field $\chi$, and also encompasses a previously studied preheating channel based upon a non-standard kinetic term.

\end{abstract}

\pacs{04.20.Fy, 98.80.Es}

\maketitle 

\section{Introduction}

Cosmology is in a new era in which it is possible to make detailed quantitative analysis for the early universe, due to the wealth of observational data stemming from important experiments such as WMAP \cite{WMAP} and the Sloan Digital Sky Survey \cite{SDSS}, amongst others. Inflation, which assumes a period of accelerated expansion after the Big Bang, is the most studied candidate to solve the monopole, horizon and planarity problems \cite{Guth,Starobinsky}, as well as explaining the anisotropies in the cosmic microwave background radiation. The simplest models of inflation yield a scale-invariant spectra of gravitational waves and energy density perturbations which act as seeds for structure formation. Most models of inflation are based on the slow-roll of scalar fields. In contrast with the cosmological constant scenario, slow-roll does not present the so-called ``exit problem'', since the period of cosmic acceleration is followed by a subsequent radiation-dominated era, along with a transient matter-dominated phase. 

Given the known equivalence between scalar models and the so-called $f(R)$ theories, where the Einstein-Hilbert action is generalized to admit a non-linear $f(R)$ term in the scalar curvature $R$, it is natural that inflation may also be obtained within the framework of the latter proposals (see Ref. \cite{review} for a thorough review). In particular, an early model for inflation relies on a quadratic addition to the linear curvature term, $f(R) = R + R^2/(6M^2)$ \cite{Starobinsky}, with WMAP normalization of the CMB temperature anisotropies indicating that  $M \sim 3\times 10^{-6}~M_P$, $M_P$ being the Planck mass \cite{WMAP5}.

A recent generalization of $f(R)$ extensions of General Relativity (GR) includes the presence of non-trivial terms in $R$ by introducing a  non-minimal coupling of the scalar curvature with matter \cite{f2}, via the action

\beq S = \int \left[ {1 \over 2}f_1(R) + f_2(R) \mathcal{L} \right] \sqrt{-g} d^4 x~~. \label{model}\eeq

\noindent This model has a rich lore of implications, both theoretical and observational. These include the deviation from geodesic motion \cite{f2}, the possibility of mimicking dark matter by leading to the flattening of the galaxy rotation curves \cite{DM} and the modelling of the accelerated expansion of the Universe at late times \cite{cosmo}. Although not related to this work, recent developments in inflationary models have led to a rehabilitation of the Higgs boson as a putative candidate for the inflaton field, provided that it is non-minimally coupled to the curvature \cite{Higgs}.

In the context of inflation, a non-minimal coupling is the key ingredient for the so-called preheating mechanism, which arises due to a scalar field $\chi$ with a variable mass term of the form $m^2_{eff} = m^2 + \xi R$; due to the oscillatory phase that follows the slow-roll regime, the dynamics of this scalar field can undergo parametric resonance, thus giving rise to the quantum production of massive particles even for relatively low values of the coupling $\xi \gtrsim 1$ \cite{preheating1,preheating2}. Furthermore, preheating may also be implemented with more evolved couplings, {\it e.g.} a quadratic coupling $R^2 \chi$ \cite{mimoso1, mimoso2} or via a non-standard kinetic term for the matter scalar field, of the form $g(R) (\partial \chi)^2 $ \cite{kinetic}.

Hence, it is quite natural to expect that the model in Eq. (\ref{model}) is capable of generalizing the preheating scenario to the case of a universal non-minimal coupling between matter and geometry. Thus, the main purpose of this work is to show that preheating indeed occurs if one assumes a generalized coupling of the form $f_2(R) {\cal L}$ in the modified Einstein-Hilbert action, instead of the {\it ad-hoc} terms $\xi R \chi^2$ or $g(R) (\partial \chi)^2 $.

However, the universality of this model implies that curvature is  coupled to all matter species, {\it i.e.} matter and radiation, besides the scalar field $\chi$. Naturally, this may potentially modify the cosmological dynamics, as given by the Friedmann equation: the latter will display terms arising not only from the Starobinsky prescription $f_1(R) = R + R^2/(6M^2)$, but also from the non-minimal coupling function $f_2(R)$. Hence, one should first establish the validity of the Starobinsky inflationary regime, that is, that these extra terms do not become dominant during the inflationary phase.

This manuscript is organized as follows: firstly, one briefly discusses the main features of the Starobinsky inflationary regime and ensuing preheating mechanism, as well as the fundamental results arising from the considered non-minimally coupled model. The third section addresses the required constraints to the non-minimal coupling that allow for Starobinsky inflation to occur. One then proceeds to the main purpose of this work and study how the non-minimal coupling enables a parametric resonance leading to preheating of the Universe. Conclusions are then put forward, and an outlook of future developments is drawn. 

Since it is not directly related to the obtained results, the analogy between the non-minimally coupled model under scrutiny and a multi-scalar-tensor theory \cite{scalar} is deferred to an appendix. A second appendix deals with the possibility of implementing an inflationary era purely via a non-minimal coupling between curvature and matter.

\section{Starobinsky inflation}

As already stated, the well known Starobinsky inflationary model \cite{Starobinsky} considers a quadratic correction to the Einstein-Hilbert Lagrangian, 

\beq f(R)  =  2\ka \left(R+{R^{2} \over 6M^2}\right)~~, \label{StarModel}\eeq

\noindent where $\ka = (16\pi G)^{-1} = M_P^2/16 \pi $ and $M \simeq 3 \times 10^{-6}~M_P$. This model can lead to an inflationary stage in the early Universe, due to the presence of the quadratic term, which ends by the dominance of the linear one.

One assumes the Robertson-Walker metric, as given by the line element

\beq ds^2 = - dt^2 + a^2(t) \left({dr^2 \over \sqrt{1-kr^2}} + d\Om^2\right)~~, \label{metric} \eeq

\noindent where $a(t)$ is the scale factor and $k$ is the spatial curvature (which is set to $k=0$ in the following calculations).

The Einstein field equations of standard $f(R)$ theories are obtained from action \eq{model}, by substituting $f_1(R) = f(R)$ and taking the trivial case $f_2(R) = 1$:

\beqa && F_1 \left(R_\mn - {1 \over 2}g_\mn R\right) =  \\ \nonumber && T_\mn + \De_\mn F_1 + {1 \over 2} (f_1 - F_1R) g_\mn ~~, \eeqa

\noindent with $F_1 \equiv f_1'(R)$ and $\De_\mn \equiv \nabla_\mu \nabla_\nu - g_\mn \square$. Introducing the Starobinsky prescription \eq{StarModel} leads to

\beqa \label{fieldeqs} && \left(1 + {R \over 3M^2} \right) \left( R_\mn - {1 \over 2} g_\mn R \right) = \\ \nonumber && {1 \over 2\ka } T_\mn + {1 \over 3M^2} \De_\mn R -{R^2 \over 12M^2} ~~.  \eeqa

\noindent  Inserting the metric \eq{metric}, these may be rewritten as

\beqa \ddot{H}-{\dot{H}^2 \over 2H}+{1\over 2} M^2 H &= & -3H \dot{H}~~, \label{StarEq1}\\ \ddot{R}+3H\dot{R}+M^2 R &= & 0~~, \label{StarEq2}\eeqa

\noindent where $H \equiv \dot{a}/a$ is the Hubble parameter, $R = 6(\dot{H} + 2H^2)$  and the dots denote time derivatives.

The cosmic acceleration can be written as $\ddot{a}/a = H^2(1-\ep)$, where $\ep \equiv -\dot{H}/H^2$ is the slow-roll parameter. In the slow-roll approximation $\epsilon \ll 1$, the former is positive defined, corresponding to an inflationary regime. Since in this approximation one also has $|\ddot{H}/(H\dot{H})| \ll 1$, the two first terms in Eq. (\ref{StarEq1}) can be neglected, thus yielding

\beq\epsilon \simeq {M^2 \over 6H^2}~~. \label{epsilon} \eeq

\noindent One may straightforwardly calculate the number of e-folds $N$, so that the scale factor increases by an amount $e^N$ during inflation. This is given by

\beq N \equiv \int_{t_i}^{t_f} H\,\mathrm{d}t \simeq {1 \over 2\epsilon_1(t_i)}~~,\label{efolds} \eeq

\noindent where $t=t_i$ and $t=t_f$ correspond to the instants of onset and end of inflation, respectively.

Inflation ends when the slow-roll parameter \eq{epsilon} becomes of order unity, as the Hubble parameter drops below $H_f \sim M/\sqrt{6}$. After this, one can no longer disregard the second time derivative of the scalar curvature in Eq. (\ref{StarEq2}). For the solution of this equation one must perform the substitution $R\rightarrow a^{-3/2} R$, 

\beq \ddot{R}+\left(M^2-{3\over 4}H^2-{3\over 2}\dot{H}\right)R=0~~. \eeq 

\noindent During reheating one has $M^2\gg \{H^2,|\dot{H}|\}$, which leads to a harmonic oscillator solution for the last equation and a damped one for the original scalar curvature, $R\propto a^{-3/2}\sin (Mt)$. In order to seek for the Hubble parameter one may neglect the {\it r.h.s.} of Eq. (\ref{StarEq1}) obtaining the solution $H(t) \propto \cos^2 (Mt/2)$, which suggests the {\it Ansatz}

\beq H(t)=f(t) \cos^2 (Mt/2)~~. \label{ReheatEq1} \eeq

\noindent Taking the solution for the slow-roll regime, $\dot{H}=-M^2/6$ and the approximation $R(t) \simeq 6\dot{H}$, one has \cite{oscilatory}

\beq R(t) \simeq 6\dot{H} =  -3M f(t) \sin [M(t-t_o)]~~, \label{ReheatEq2} \eeq

\noindent where $t_o$ denotes the onset of the oscillatory regime and 

\beq f(t) = \left[ {3 \over M} + {3 \over 4}(t-t_o) + {3 \over 4M} \sin (M [t-t_o]) \right]^{-1}  ~~. \eeq

In the $M(t-t_o) \gg 1 $ regime one can consider the approximation 

\beq f(t) \simeq {4 \over 3 (t-t_o)}~~, \eeq

\noindent for this last set of equations. This yields

\beq H(t) \simeq -{4 \over 3(t-t_o)} \cos^2 \left[{M\over 2} (t-t_o) \right] ~~, \label{Hlate} \eeq

\noindent so that $\left< H \right> \simeq (2/3) (t-t_o)^{-1}$ (indicating that during the Starobinsky reheating phase, the Universe behaves as if matter-dominated) and 

\beq R(t)  \simeq - {4M \over t-t_o} \sin [M(t-t_o)] ~~. \label{Rlate} \eeq

\subsection{Standard preheating}

Almost all the matter that constitutes the Universe, in the subsequent radiation-dominated era, was created during the reheating process at the end of the inflation through the gravitational particle creation, which occurs via oscillations of the Ricci scalar.

This particle production can be effectively described by scalar fields\footnote{For a discussion involving two scalar fields see {\it e.g.} Ref. \cite{OBR87}.}. For simplicity, one considers a scalar field $\chi$ with mass $m$, along with a non-minimal coupling with the scalar curvature,

\beqa \label{actionpre} S &=& \int \sqrt{-g}{\rm d}^4x \times \\ \nonumber && \left[ {f(R)\over 2\kappa^2}-{1\over 2} g^{\mu \nu} \partial_{\mu} \chi \partial_{\nu} \chi-{1\over 2} m^2 \chi^2-{1\over 2} \xi R \chi^2 \right] ~~, \eeqa

\noindent where $f(R)$ is equivalent to the function $f_1(R)$ in the action of our model. The variation of the action with respect to the field $\chi$ leads to 

\beq \square \chi - m^2 \chi -\xi R \chi=0~~. \eeq

\noindent This quantum field $\chi$ can be decomposed in modes as depicted below

\beqa \label{QuantDec} \chi(t,{\bf x}) &=& {1 \over (2\pi)^{3/2}} \int d^3k \times \\ \nonumber 
&&\left[ a_k \chi_k(t) e^{-i {\bf k} \cdot {\bf x}} + a^\dagger_k \chi^*_k(t) e^{i {\bf k} \cdot {\bf x}} \right] ~~,  \eeqa

\noindent where one has the creation $a^\dagger_k$ and annihilation $a_k$ operators of particles with mass $m$ and momentum ${\bf k}$.

The field $\chi$ can be quantized in curved spacetime by generalizing the basic formalism of quantum field theory in flat spacetime. Then each Fourier mode $\chi_k (t)$ obeys the following equation of motion 

\beq \ddot{\chi}_k+3H \dot{\chi}_k+\left( {k^2\over a^2}+m^2+\xi R \right) \chi_k=0~~, \label{QuantMod}\eeq

\noindent where $k=|\bf{k} |$ is the comoving wavenumber. After the substitution $\chi_k\rightarrow a^{-3/2}\chi_k$, this last equation becomes

\beq \ddot{\chi}_k+\left( {k^2\over a^2}+m^2+\xi R-{9\over 4} H^2 -{3\over 2} \dot{H} \right)\chi_k=0~~. \label{QuantModNew}\eeq

The last two terms in Eq. (\ref{QuantModNew}) can be neglected if $|\xi|>1$. Taking the approximation of the scalar curvature given by Eq.~(\ref{Rlate}) in the regime $M (t-t_{{\rm o}}) \gg 1$, the same equation becomes 

\beq \ddot{\chi}_k+\left[ {k^2\over a^2}+m^2-{4M \xi\over t-t_{{\rm o}}} \sin \{ M (t-t_{{\rm o}}) \} \right]\chi_k \simeq 0~~.\label{Xeq2} \eeq

\noindent Defining the varying frequency as

\beq\omega_k^2\equiv{k^2\over a^2}+m^2-{4M\xi\over (t-t_{\mathrm{o}})} \sin\{ M(t-t_{\mathrm{o}}) \}~~,\eeq

\noindent the same equation can be written as that of a parametric oscillator $\ddot{\chi}_k+\omega_k^2\chi_k \simeq 0$.

The particle production is achieved via the oscillating term in the above expression, through the already mentioned parametric resonance. Inserting a variable $z$ constructed from the relation $M(t-t_{\mathrm{o}})=2z \pm \pi/2$, where the different signs correspond to the sign of $\xi$, leads to the Mathieu equation 

\beq {\mathrm{d}^2\chi_k\over \mathrm{d}z^2}+\left[ A_k-2q \cos (2z) \right]\chi_k \simeq 0~~, \label{mathieu0} \eeq

\noindent where the parameters $A_k$ and $q$ determine the strength of parametric resonance and are given by

\beq A_k={4k^2\over a^2M^2}+{4m^2\over M^2}\,,\qquad q={8|\xi|\over M (t-t_{{\rm o}})}~~. \label{defsmathieu} \eeq

\noindent The amount of parametric resonance can be described by a stability-instability map of the Mathieu equation~\cite{Mathieu,kofman1,preheating2}, which have instability bands where the perturbations grow exponentially with different rates~\cite{kofman1,kofman2}.

In an expanding universe both $A_k$ and $q$ vary in time. Since the field passes many instability and stability bands, the growth changes with the cosmic expansion. The non-adiabaticity of the change of the frequency $\omega_k$ can be estimated by the quantity $r_{\mathrm{na}} \equiv \left| \frac{\dot{\omega}_k}{\omega_k^2} \right|$ which, for small $k$ and $m$, becomes

\beq 
r_{\rm na}\simeq M{| \cos\{ M(t-t_{\mathrm{o}}) \}/(t-t_{\mathrm{o}})|\over |4M\xi \sin\{ M(t-t_{\mathrm{o}}) \}/(t-t_{\mathrm{o}})|^{3/2}}~~.\eeq

\noindent The non-adiabatic regime corresponds to $r_{\mathrm{na}} \gtrsim 1$. However, as one can see, the condition $M(t-t_{\mathrm{o}})=n\pi$ leads to a more efficient non-adiabatic particle production, since $r_{\mathrm{na}} \gg 1$, and this occurs when the Ricci scalar vanishes.

For instance, in the Starobinsky model~(\ref{StarModel}) the massless $\chi$ particles are resonantly amplified for $|\xi| \gtrsim 3$ \cite{preheating2}, and the massive ones with $m \sim M$ can be created for $|\xi| \gtrsim 10$. These computations can be further extended to take into account non-linear effects, like the coupling between different modes, which can be important at the end of the preheating stage \cite{kofman2,Khle97}; the backreaction of the Ricci scalar that emerges from the additional contribution of the energy density of the created particles to the background cosmological dynamics via the Friedmann equation can also be considered, but with no significant impact on the reheating process \cite{preheating2}. 

One remarks that preheating is not a mandatory ingredient of inflationary models: indeed, in the standard reheating scenario with $\xi = 0$, particle production is still possible, albeit it is achieved via a perturbative decay of the inflaton degree of freedom ({\it vis-\`{a}-vis} the curvature, as discussed in the previous section and Ref. \cite{analogy}) into two $\chi$ quanta \cite{kofman1,kofman2,preheating0}. 

From \eq{defsmathieu}, one sees that $\xi = 0 $ implies $ q = 0$, so that \eq{mathieu0} describes a parametric oscillator driven by the expansion of the Universe alone; hence, the aforementioned crossing of the broad parametric ressonance bands does not occur, rendering standard reheating rather inefficient when compared with preheating (see Ref. \cite{scalaron} for a recent discussion of the dynamics of thermalization).

\subsection{Non-minimal coupling}

One addresses here a model that exhibits a non-minimal coupling between geometry and matter, as expressed in the action \eq{model}. Varying with respect to the metric leads to the modified field equations

\beqa \label{modfieldeqs} && \left(F_1 - 2 F_2 \rho \right) \left(R_\mn - {1 \over 2}g_\mn R\right) = f_2 T_\mn + \\ \nonumber &&  {1 \over 2} (f_1 - F_1R) g_\mn + F_2 \rho R g_\mn + \De_\mn \left(F_1 - 2F_2 \rho\right)~~, \eeqa

\noindent with $F_i \equiv f_i'(R)$; inserting the Robertson-Walker metric \eq{metric}, the temporal and spatial components read

\beqa && 3\left(F_{1}-2F_{2}\rho\right)H^{2} = (1 + f_2) \rho  \\ \nonumber && -{1\over 2}\left[f_{1}-\left(F_{1}-2F_{2}\rho\right)R\right] -3H\partial_{0}\left(F_{1}-2F_{2}\rho\right)~~,\eeqa

\noindent and

\beqa && -2\left(F_{1}-2F_{2}\rho\right)\dot{H} = \left(1+f_{2}\right)\left(\rho+p\right)+  \\ \nonumber && \left(\partial_{0}\partial_{0}-H\partial_{0}\right)\left(F_{1}-2F_{2}\rho\right) ~~.
\eeqa

\noindent As expected, GR is recovered by setting $f_1(R) = 2\ka R $ and $f_2(R) = 1$.

Resorting to the Bianchi identities, one concludes that the energy-momentum tensor may not be conserved in a covariant way

\beq \nabla^\mu T_{\mu\nu}={F_2 \over f_2}\left(g_{\mu\nu}{\cal L}-T_{\mu\nu}\nabla^\mu R\right)~~. \label{cov} \eeq

\noindent In fact, as expected in the absence of the coupling, $f_2(R)=1$, the covariant energy-momentum conservation is recovered. This feature implies that the motion of the matter distribution described by a Lagrangian density ${\cal L}$ does not follow a geodesic curve. Clearly, a violation of the Equivalence Principle may emerge if the {\it r.h.s.} of the last equation varies significantly for different matter distributions, which suggests a method of testing the model and imposing constraints to the associated coupling constants. Of course, this putative violation is detectable only in astrophysical and/or cosmological contexts.

\section{Inflationary regime}
\label{section3}

\subsection{Non-conservation of energy-momentum tensor}

In order to extend the standard preheating scenario in the context of Starobinsky inflation, one assumes a linear coupling between matter and geometry \footnote{Although not discussed here, it can be hinted from \eq{fourier} that higher order additions $f_2(R) = 1+ \xi(R/M^2)^n$, with $n> 1$ would (in the $f_2(R) \approx 1$ regime) deviate the dynamics from the desired parametric resonance --- although a more convoluted preheating may perhaps still be achieved.}, as well as the usual Starobinsky curvature term,

\beqa \label{ffunc} f_1(R) &=& 2\ka \left(R + {R^2 \over 6M^2}\right)~~, \\ \nonumber f_2(R) & =& 1 + 2 \xi {R \over  M^2}  ~~,\eeqa

\noindent where $\xi$ is a dimensionless parameter of the model and $M$ has dimensions of mass.

One considers that matter is described by a perfect fluid with the appropriate energy-momentum tensor,

\beqa \label{defSET} T_{\mu \nu}=-{2 \over \sqrt{-g}}{\delta(\sqrt{-g}{\cal L}_m)\over \delta(g^{\mu\nu})} = \\  \nonumber \left( \rho +p\right) U_{\mu }U_{\nu }+pg_{\mu
\nu } ~~,\eeqa

\noindent where $U_\mu$ is the four-velocity, $\rho$ is the density and $p$ is the pressure, which is related with the former by the equation of state (EOS) $p = \om \rho$.

Three different matter species are assumed, for completeness: radiation, with ${\cal L}_r = p_r$ and $\om_r = 1/3$, matter (taken as an ultrarelativistic particle-antiparticle plasma) with ${\cal L}_m = -\rho_m$ \cite{fluid}  and $\om_m = 1/3$ and a scalar field $\chi$ with 

\beq {\cal L}_\chi = -{1 \over 2} \partial_\mu \chi \partial^\mu \chi - V(\chi)~~.\eeq

In order to obtain the Friedmann equation, one requires the computation of $\dot{\cal L}_i$, obtained from the time component of the non-conservation law for the energy-momentum tensor of each species, as depicted below.

Defining the energy density $\rho_\chi = -\partial_\mu \chi\partial^\mu \chi / 2 + V(\chi)$ and pressure $p_\chi = {\cal L}_\chi$ of the scalar field $\chi$, one gets a formally equivalent result for this component and for radiation,

\beqa \dot{\rho}_j &=& -\left( 3H + {F_2 \over f_2} \dot{R} \right) (\rho_j + p_j) =  \label{cons} \\ && -\left( 3H + {\dot{R} \over R} \right) (\rho_j + p_j) \rightarrow \nonumber \\ && \cases{ \dot{\rho}_r = -\left(4H + {4 \over 3 } {\dot{R} \over R} \right) \rho_r \cr ~\cr \ddot{\chi} = - \left(3H + {\dot{R} \over R} \right) \dot{\chi}  - V'(\chi)  } ~~. \label{cons2} \eeqa

\noindent assuming that the curvature is high enough so that $R > M^2 / (2 \xi) \rightarrow f_2 (R) \approx 2\xi R /M^2$.

As has been seen in the context of the accelerated expansion of the Universe \cite{cosmo}, one finds that the relativistic matter component is covariantly conserved,

\beqa \nabla^\mu T^m_{\mu 0} &=& {F_2 \over f_2}\left(T^m_{\mu 0 } - g_{\mu 0 } {\cal L}_m\right) \nabla^\mu R = \\ \nonumber  && - {F_2 \over f_2}\left(T^m_{0 0 }+ g_{0 0 } \rho_m\right) \dot{R} =0~~,  \eeqa

\noindent and hence,

\beqa \label{consrhom} \dot{\rho}_m& =& -3H(\rho_m + p_m) = -4 H \rho_m \rightarrow \\ \nonumber  \rho_m(t) &=& \rho_{mi} \left({a_i \over a}\right)^4~~.  \eeqa

In the case of radiation, one may integrate directly the above expression to obtain the exact result

\beqa \label{anal} && \rho_r(t) = \rho_{ri} \left({a_i \over a(t)}\right)^4 \left[{f_2(R_i) \over f_2(R)}\right]^{4/3} = \\ \nonumber && \rho_{ri} \left({a_i \over a(t)}\right)^4 \left[{M^2 + 2 \xi R_i  \over M^2 + 2\xi R }\right]^{4/3} \approx  \rho_{ri} \left({a_i \over a}\right)^4 \left({R_i \over R}\right)^{4/3} ~~,  \eeqa 

\noindent where $\rho_{ri}$, $a_i$ and $R_i$ are initial values of the radiation density, scale factor and curvature, respectively, and the final step assumes that $f_2(R) \gg 1 \rightarrow R \gg M^2/2\xi$; conversely, in the weak coupling regime $f_2(R) \sim 1 \rightarrow R \ll M^2/2\xi$ the above equation naturally reduces to the conservation law $\rho \propto a^{-4}$.

Although both the matter field $\chi$ and radiation admit the same Lagrangian density, ${\cal L} = p$, the former is not characterized by a constant EOS parameter $\om_\chi = p(\chi) / \rho(\chi)$ --- except for massless, non-self-interacting particles, $ V (\chi) = 0$, where $p_\chi = \rho_\chi = \dot{\chi}^2/2 \rightarrow \om_\chi = 1$, such that the scalar field behaves as ultra-stiff matter. One may compute the evolution of  $\rho_\chi$ in the latter case, following \eq{cons2} for $\ddot{\chi}$:

\beqa \label{consdenschi} \rho_\chi(t) =  \rho_{\chi i} \left({a_i \over a(t)}\right)^6 \left[{f_2(R_i) \over f_2(R)}\right]^2  & = & \\ \nonumber  \rho_{\chi i} \left({a_i \over a}\right)^6 \left({R_i \over R}\right)^2 ~~,  \eeqa

\noindent for an initial value of the scalar field density $\rho_{\chi i}$.

The dissimilitude between Eqs. (\ref{consrhom}) and (\ref{anal}) is striking, since in GR one naturally finds that matter follows the same conservation law as radiation, as they share the common EOS $p = \rho/3$. However, one cannot extend this common behaviour into the model here considered, {\it i.e.} assume that both components will obey the same non-conservation law.

Indeed, one of the major features of the model embodied in action \eq{model} is that the Lagrangean density itself acquires a direct physical significance, which appears explicitly in the modified field Eqs. (\ref{modfieldeqs}) and the covariant non-conservation law \eq{cov} --- instead of only implicitly, via the definition of the energy-momentum tensor. Hence, although the EOS of matter and radiation is the same, they follow distinct temporal evolutions because of the different Lagrangian densities, $ {\cal L}_r = p $ and ${\cal L}_m = \rho$.

This said, one remarks that the choice of Lagrangian density for matter is the subject of some debate in the literature, mostly because it is never directly involved in computations in GR (see Ref. \cite{fluid} and references therein). However, this discussion is not crucial to the results here presented, nor is the prescription ${\cal L}_m = \rho$ paramount to their derivation: if one chose ${\cal L}_m = {\cal L}_r = p$ instead, the non-conservation laws for matter and radiation would obviously be the same --- but this would only alter the numerical coefficients affecting the {\it r.h.s.} of \eq{modFriedmann}, not the ensuing discussion concerning the validity of the Starobinsky regime; from a less technical standpoint, one could invoke to dominance of radiation in the primitive Universe to ascertain the lack of importance of the choice for ${\cal L}_m$ in the present context.

Given the condition $f_2(R) \gg 1 \rightarrow R \gg M^2 / 2\xi$, one sees that the value of the ``mass-scaling'' parameter $\xi$ does not play a role in the above non-conservation laws. Furthermore, since the slow-roll regime yields $\dot{R} = -24H^4 \ep$, with $\ep \ll 1$, one does not expect the non-trivial contributions to make a significant impact, so that the three matter species evolve in a classical way. Finally, given the purpose of generalizing the preheating mechanism in the context of the non-minimal coupling model, one assumes that $\xi > 1$, so that the condition $R \gg  M^2 / (2 \xi) $ is always fulfilled during slow-roll (which ends when $R \sim M^2$), and so the above non-conservation laws hold throughout this phase.

\subsection{Starobinsky regime}

One aims to obtain a regime of Starobinsky inflation, where the terms derived from the curvature contribution $f_1(R) = 2\ka \left(R+ R^2/(6M^2)\right)$ dominate, thus leading to the inflationary evolution described in the previous section. For this, one should rederive the Friedmann equation, taking into consideration also the effect of the non-minimal coupling $f_2(R) = 1 + 2\xi R/M^2$, and set the conditions which should be fulfilled so that these are subdominant with respect to the latter.

This said, using the $0-0$ component of the Einstein equations and the (non-)conservation laws obtained above, one gets the modified Friedmann equation,

\beqa \label{modFriedmann} - \left[ 1 - {2\xi \over \ka M^2} \left( 3 \rho_m - {11 \over 3} \rho_r - {9 \over 2} \dot{\chi}^2 -V \right) \right] H^2 =  \\ \nonumber  {1 \over M^2}\left[ {R \over 6} + {2\xi \over \ka } \left( {4 \over 3} \rho_r + {3 \over 2} \dot{\chi}^2 \right) \right] \dot{H} + \\ \nonumber  {1 \over 3M^2} \left[ {2\xi \over \ka R} \left(  {4 \over 3}\rho_r +  \dot{\chi}^2 \right) -1  \right] H\dot{R} + {\rho \over 6\ka}~~,   \eeqa

\noindent where $\rho = \rho_r + \rho_m + \rho_\chi$ is the total energy density.

One assumes that the scalar field $\chi$ has negligible initial potential $V(\chi)$ and kinetic energy $\dot{\chi}^2$, so that one may disregard it in the above equation; also, one assumes that the values for these quantities at the end of the slow-roll regime are still well below the dominant contribution, {\it i.e.} the scalar field will only play a relevant role in the (p)reheating phase. Recalling that  $R \gtrsim M^2$ in the Starobinsky regime, the  remaining conditions for Starobinsky inflation amount to 

\beq {2\xi \rho_j \over \ka } \ll M^2 \lesssim R ~~,  \label{cond} \eeq

\noindent with $j = r,~m$. Since ultrarelativistic matter behaves as radiation, one may use the Stefan-Boltzmann law $\rho_j \propto g_j T^4 $ (where $g_j$ is the number of degrees of freedom and $T $ is the temperature, assuming thermal equilibrium) to write

\beq {T \over M_P } \ll {1 \over (2 \xi)^{1/4}} \sqrt{M \over M_P} \approx 10^{-3} \xi^{-1/4}~~, \label{condT}\eeq

\noindent after using that $M \approx 3 \times 10^{-6}~M_P$ \cite{WMAP5}.

As stated before, one assumes that $\xi > 1$, so that this parameter does not play a role in the evolution of matter during slow-roll, but it becomes relevant in preheating, at the onset of the condition $R \sim M^2 / (2 \xi)$. This, coupled with the expected smallness of the $\dot{R}/R$ terms in \eq{cons2}, leads one to consider that the temporal evolution of the density of the considered matter species does not differ significantly from that expected in standard Starobinsky inflation. Indeed, using the slow-roll approximation

\beq R = 6H^2(2-\ep) \rightarrow \dot{R} \simeq -24 H^3 \ep ~~, \eeq

\noindent one gets 

\beq \dot{R}\simeq -2 HR \ep ~~,  \eeq

\noindent so that 

\beqa \dot{\rho}_r &=& -{4 \over 3} \left(3  - 2 \ep \right)  H \rho ~~, \label{consradslow} \\  \ddot{\chi} &=& - \left (3 -2 \ep\right) H \dot{\chi} - V'(\chi) ~~, \label{conschislow} \eeqa

\noindent and, since $\ep \ll 1$, the contribution from the non-minimal coupling may be disregarded, and the energy-momentum tensor of all matter species is almost completely conserved (see paragraph \ref{slow-cooling} for more detail). In particular, this means that no decreasing effect on the cooling rate is attained, {\it i.e.} the Hubble parameter always dominates the {\it r.h.s.} of the considered equation and the density (and temperature) never rises due to the non-minimal coupling. Hence, it suffices to evaluate Eqs. (\ref{cond}) and (\ref{condT}) at their maximum initial values.

Upper bounds for $\xi$ can be obtained by assuming that the initial inflationary temperature is, at least, of the grand unified theory (GUT) scale $T_{GUT} \sim 10^{15} ~GeV \sim 10^{-4}~M_P$. Hence, one obtains 

\beq \xi \ll 10^{-12} \left({M_P \over T_{GUT} }\right)^4 \sim 10^{4} ~~. \label{consbe} \eeq

\noindent The obtained range $ 1 < \xi < 10^4 $ can be translated into an equivalent mass scale appearing in the non-minimal coupling term $f_2(R) = 1 + R/M_2$, with $M_2 = M /\sqrt{2\xi} \sim 10^{-2}M-M$; hence, one does not introduce a strong mass hierarchy in the non-trivial components of the models, and may consider only the mass scale $M$ as the key ingredient to the model.

Finally, one sees from \eq{condT} that the maximum initial temperature compatible with $\xi > 1$ is given by $T_{max} \sim 10^{-3}~M_P \sim 10~T_{GUT}$; hence, our model seems to be particularly well suited for a GUT scale inflation, although it fails for Planck scale inflation. The latter, in turn, is not quite desirable on account of most of the baryogenesis mechanisms (see however, Ref. \cite{BCKP}).

The above constraint is not regarded as a caveat of the present approach, since it is not in contradiction with the required minimum number of e-folds $ N > 60$: indeed, inserting $H_i \sim T \sim T_{GUT} $ into the result from the Starobinsky regime $N \simeq 3(H_i / M)^2$, one obtains $N = 10^3 - 10^5 > 60$, having again used the CMB value $M \approx 3\times 10^{-6}~M_P$.

\subsection{Slow-cooling due to the non-minimal coupling}
\label{slow-cooling}

As argued before, it is not feasible to use the effect of the non-minimal coupling to alleviate the supercooling ensued after inflation. The impossibility of attaining this ``slow-cooling'' regime is assessed by simply writing the final radiation density after slow-roll, as obtained from the direct analytical solution \eq{anal}:

\beq \rho_{rf} =  \rho_{ri} \left({a_i \over a_f}\right)^4 \left({R_i \over R_f}\right)^{4/3}= \rho_{ri} e^{-4N} (4N)^{4/3} ~~, \eeq 

\noindent having used $R_f = 6H_f^2 = M^2$ and $R_i \simeq 12H_i^2 \simeq 4NM^2 =4 N R_f$. Hence, one may only obtain an increase in density by a factor $(4N)^{4/3}$ (when compared with the value obtained without the effect of the non-minimal coupling) or, using the Stefan-Boltzmann law $\rho_r \propto T^4$, a factor $(4N)^{1/3}$ in temperature. Considering that the number of e-folds is constrained to be of the order $10^3-10^5$, due to the validity conditions of the Starobinsky regime already discussed, one obtains a compensating factor of order $10-100$, manifestly insufficient to end inflation with a ``warm'' Universe. Thus, one must deal with the issue of implementing reheating after slow-roll, as will be considered in the following section.

\section{Reheating}

\subsection{Starobinsky evolution}
Assuming that the condition \eq{cond} holds, one may approximate the modified Friedmann equation (\ref{modFriedmann}) by the one arising in standard Starobinsky inflation, so that the slow-roll treatment gives rise to the temporal evolution already discussed.

Given that $ \xi > 1 $, the end of slow-roll $\ep \sim 1$ corresponds to $R \sim M^2 > M^2 / (2 \xi) $; hence, one might assume that the non-minimal coupling $f_2(R) = 1 + 2\xi R/ M^2$ would take hold of the dynamics until it too gets close to unity, when $R < M^2 / (2 \xi)$. However, one should keep in mind that its effect is mediated by the value of the density of the considered matter species, according to \eq{cond}; since this condition stems from the modified Friedmann equation, Eq. (\ref{modFriedmann}), regardless of the validity of the slow-roll condition $\ep \ll 1$, and that the density has dropped exponentially during inflation. Hence, one concludes that the post-inflationary dynamics is achieved by disregarding the terms due to the non-minimal coupling: the evolution of the Hubble parameter and the curvature is then given by Starobinsky model post-inflationary phase described by Eqs. (\ref{ReheatEq1}) and (\ref{ReheatEq2}).

\subsection{$f_2 \approx 1$ regime}

As stated before, one attempts now to use the dynamics of the matter scalar field $\chi$ to provide the required reheating of the Universe, which after inflation is essentially at zero temperature. For concreteness, one dubs the timespan during which $f_2(R) \approx 2\xi R/ M^2$ as the {\it coupled} regime, while $f_2(R) \approx 1 $ is naturally termed as {\it uncoupled}.

Clearly, since the curvature oscillates, one cannot assume that the coupled regime prevails at the onset of oscillation, as during each of these the curvature vanishes and one has $f_2(R) = 1$. Furthermore, since its amplitude also drops with $t^{-1}$, at $t = t_o + 2\xi /M $ (that is, after $ \xi/\pi $ oscillations) it is below $M_2^2 \equiv M^2/ 2\xi $, a point at which one can no longer assume that $f_2(R) \approx R/M_2^2 \gg 1$.

From \eq{consbe}, the number of oscillations before one crosses over to the $f_2(R) \approx 1$ regime is constrained to be less than $\sim 10^3$; if $\xi $ is close to unity (and no hierarchy exists between $M$ and $M_2$), not even one oscillation is completed. Bearing this in mind, one considers only this uncoupled regime; this can be considered a worst case scenario, given that one neglects any reheating that might occur during the coupled regime. A numerical analysis (not performed here) should allow for a better discrimination of both regimes and assess the impact on the dynamics and reheating.

However, disregarding this ``coupled reheating'' should bear no relevance on the final result, since the reheating generated during the uncoupled regime should be much more important: indeed, this arises from the term $F_2/f_2$ appearing in \eq{cons2}, which (since $F_2 = \mathrm{const.}$) rises as the curvature falls and $f_2(R)$ drops to unity.

As a result of the above discussion, one proceeds to consider the $f_2(R) \approx 1$ uncoupled late-time regime as the only channel to implement a suitable reheating mechanism in the model under scrutiny. Naturally, this does not coincide with the minimally coupled regime $f_2(R) = 1$, since the kinetic term $F_2$ is non-vanishing --- only the covariant non-conservation laws for matter and radiation collapse to their GR counterparts, as \eq{consrhom} becomes identical to \eq{anal}, {\it i.e.} $\rho_m \propto \rho_r \propto a^{-4}$.

One begins by noticing that \eq{cons2} no longer applies, since $F_2/f_2 \rightarrow 2\xi / M^2$. As before, one may resort to the analytical integration of the non-conservation law for the radiation pressure or a massless scalar field with constant EOS parameter $\om_\chi = 1$, \eq{anal}:

\beqa \rho_j(t) &=& \rho_{2j} \left({a(t_2) \over a(t) }\right)^{3(1+\om_j)} \left[{f_2(R_2) \over f_2(R)}\right]^{1+\om_j} = \\ \nonumber  && \rho_{0j} \left({a(t_2) \over a(t) }\right)^{3(1+\om_j)} ~~,\eeqa

\noindent where $t=t_2$ refers to a moment at the onset of the regime $R \ll M^2 / (2\xi) \rightarrow f_2(R) \approx 1$. One obtains the standard GR density evolution of a fluid with constant EOS parameter $\om_j$, and thus one concludes that no reheating occurs due to the non-minimal coupling between curvature and matter (which is always covariantly conserved), irrespectively whether radiation or a massless scalar field.

\subsection{Preheating}
One is left with the possibility of a non-vanishing potential $V(\chi)$; one assumes that $V(\chi) = m^2 \chi^2$, so that $\chi$ is a quantum field which may be decomposed (in the Heisenberg representation) as in \eq{QuantDec}, such that each Fourier mode $\chi_k$ obeys the differential equation
 
\beq \ddot{\chi}_k + \left( 3H + {F_2 \over f_2} \dot{R}\right)  \dot{\chi}_k + \left({k^2 \over a^2} + m^2 \right) \chi_k = 0~~. \eeq

Introducing the field redefinition $X_k \equiv a^{3/2} f_2^{1/2} \chi_k$ and the new variable $2z = M(t-t_o) \pm \pi/2$ (depending on the sign of $\xi$), one can absorb the $\chi'_k$ term, obtaining

\beqa \label{fourier} X''_k + \bigg[\left({ 2 k \over a M}\right)^2 + \left({2m \over M}\right)^2 - 3 {H' \over M} - 9 {H^2 \over M^2} + \\ \nonumber {1 \over 2} {F_2 \over f_2} \left( {1 \over 2} {F_2 \over f_2} R'^2 -6 {H R' \over M}- R'' \right) \bigg] X_k = 0~~,   \eeqa

\noindent with the prime denoting differentiation with respect to $z$. Notice that the redefined field $X_k$ closely approaches the one obtained via the  transformation performed in standard preheating scenarios, $\chi_k \rightarrow \chi_{0k} \equiv a^{3/2} \chi_k $ as the curvature drops and $f_2(R) \rightarrow 1$.

Moreover, one could choose to perform the latter transformation, thus absorbing only the term $3H \dot{\chi}_k$. The result would have yielded a similar equation to \eq{fourier}, but with an added $\chi'_k$ term involving the non-minimal coupling factor $F_2/f_2$. A similar consideration was put forward in another study, in the context of reheating via a matter scalar field with a non-standard kinetic term \cite{kinetic}.

The generalization is obvious: while standard preheating resorts to a non-minimal coupling in the mass term only (which enables an oscillating effective mass term $m^2_{eff} = m^2 + \xi R$) and the latter study considers a non-minimal coupling to the kinetic term only, our proposal does not distinguish between the components of the Lagrangean density (or, in fact, between ${\cal L}_j$ for different matter species).

In the $f_2(R) \approx 1$ regime, \eq{fourier} reads

\beqa \label{equation1} X''_k + \bigg[\left({ 2 k \over a M}\right)^2 + \left({2m \over M}\right)^2 - 3 {H' \over M} - 9 {H^2 \over M^2} + \\ \nonumber {\xi \over M^2} \left( \xi {R'^2 \over M^2} - 6 {H R' \over M}- R'' \right) \bigg] X_k = 0~~. \label{hilleq} \eeqa

In the late-time approximation $z \gg 1$, one can write

\beqa H(z) \simeq {M\over 3 z} \left[ 1 + \sin (2z) \right] ~~, \\ \nonumber R(z) \simeq 3M H'(z) \simeq {2M^2 \over z} \cos(2z) ~~, \eeqa

\noindent so that the terms due to the Hubble expansion inside the brackets of \eq{fourier} read

\beqa 3{H' \over M} & = &{2 \over z} \cos(2z)~~, \\  \nonumber 9{H^2 \over M^2} & = &\left({ 1 + \sin(2z) \over z}\right)^2 ~~, \eeqa

\noindent and $H' \gg H^2/M$, as expected. The terms arising from the non-minimal coupling become

\beqa \label{Rpp} \left(\xi {R' \over M^2}\right)^2 &=& {16 \xi^2 \over z^2} \sin^2(2z)~~, \\ \nonumber 6 \xi {H R' \over M^3} &=& - {8 \xi \over z^2} \sin(2z) [1+\sin(2z)] ~~, \\  \nonumber \xi { R'' \over M^2} &=& -{8 \xi \over z} \cos(2z)~~.  \eeqa

Since the uncoupled regime implies $R < M^2 / (2\xi) \rightarrow z/(2\xi) > 1$, one concludes that the last term $\xi R''/ M^2$ dominates the other two arising from the non-minimal coupling (although different in scope, a discussion concerning the composite effect of different harmonics in the Mathieu \eq{hilleq} can be found in Ref. \cite{mimoso2}). Furthermore, one sees that the latter compares with the dominant term from the Hubble expansion as

\beq \xi {R'' \over M^2} = -4 \xi{3H' \over M} ~~,\eeq

\noindent indicating that the non-minimal coupling dominates the dynamics for $\xi > 1 $. Hence, \eq{equation1} simplifies to

\beq X''_k + \left[\left({ 2 k \over a M}\right)^2 + \left({2m \over M}\right)^2 - \xi {R'' \over M^2} \right] X_k = 0~~, \label{equation2}  \eeq

\noindent which, inserting \eq{Rpp}, can be cast as a Mathieu equation,

\beq X''_k + \left[A_k - 2q \cos(2z) \right] X_k = 0~~, \label{mathieu} \eeq

\noindent with

\beq A_k = \left({2k \over a M}\right)^2 + \left(2 m \over M\right)^2 ~~~~, ~~~~ q = {4 \xi \over z}~~.\eeq

This is exactly the same equation as in the usual preheating scenario obtained via an {\it ad-hoc} coupling term of the form $\xi R\chi^2$. Thus, one may extrapolate directly from the results already available for this reheating model. In particular, one recalls that massless particles may be produced for values of the coupling parameter as low as $\xi \gtrsim 3$, and massive particles for $\xi \gtrsim 10$. Hence, the allowed range $ 1 < \xi < 10^4$ enables an ample production of massive particles.

\section{Discussion and Conclusions}

In this work one demonstrates that a generalized coupling between matter and geometry naturally extends the preheating mechanism in the context of the Starobinsky inflationary model. This novel preheating scenarios arises from a scalar field $\chi$ with a scalar curvature-dependent effective mass term $m_{eff}^2 = m^2 + \xi R$ or with a non-canonical kinetic term. One finds that a generalized coupling of the form $f_2(R) {\cal L}$ (with a linear $f_2(R) = 1 + 2\xi R/M^2$) leads to a dominant term of the form $\ddot{R}/M^2$ in the Mathieu equation that describes the quantum production of particles with momentum $k$; given that $R(t)$ is rapidly oscillating in the reheating era, one gets $\ddot{R}/M^2 \sim R$, and the aforementioned preheating mechanism is fully recovered.

The generalized coupling implies that one could also consider a possible dominance of the terms arising from $f_2$ in the modified Friedmann equation, since these are coupled not only with the pressure and density of the scalar field $\chi$, but also with matter and radiation. One finds that constraining the dynamics so that a Starobinsky inflationary era occurs translates into a mild condition for the coupling parameter $1 < \xi < 10^4$, more than sufficient to allow for an efficient production of massive particles, thus leading to a successful reheating of the Universe.

The obtained result should not be viewed as a sole example of the explanatory power of the generalized non-minimal coupling model under scrutiny, but regarded together with other available results: namely, the description of the galaxy rotation curves via a mimicking ``dark matter'' mechanism \cite{DM}, and the possibility of accounting for the current accelerated expansion of the Universe \cite{cosmo}. As discussed in those works, the lower values of $R$ at the late-time and large-distance regimes where these phenomena arise, respectively, naturally lead to the consideration of an inverse power-law coupling $f_2(R) \approx R^{-n}$, with $n > 0$.

On the contrary, considerations regarding inflation must be implemented via a direct power-law coupling (in this work, a linear one), which guarantee that its effect does not grow after reheating, leading the Universe away from the desired radiation-dominated era.

Conversely, the inverse power-laws considered in Refs. \cite{DM,cosmo} bear no significance in the present context, given that the curvature was sufficiently high during the inflationary era to render its effects completely negligible.

In order to unify these different couplings, one should consider that each regime (early {\it vs.} late time, central {\it vs.} long range) is dominated by the influence of a particular contribution to an overall, yet unknown non-minimal coupling function, which can then be written as a Laurent series,

\beq f_2(R) =  \sum_{n = - \infty}^{\infty} \left({R \over M_n^2}\right)^n ~~. \label{Laurent} \eeq

Naturally, other particular phenomena and environments may be of interest, if the typical values of the curvature are such that yet unprobed terms of the above series dominate the dynamics.

Finally, one discusses another context where the influence of a linear coupling $f_2(R)$ may be addressed: the central regions of a spherical body (such as a star), as its core might be sufficiently dense to constraint its effect \cite{hydro}.

Using the hydrostatic equilibrium of the Sun as a test case, it was found that the central density and available observational precision of central observables yield only a modest bound for the non-minimal coupling strength; translated into the notation of this work, one gets $\xi \ll 10^{78}$, several orders of magnitude above the upper limit of the range obtained here, $1 < \xi < 10^4$.

With the above discussion in mind, the current work may be regarded as yet another effort to reconstruct the full form (\ref{Laurent}) of the non-minimal coupling --- in this case by constraining its linear term in a far more effective fashion than through the aforementioned study of solar hydrostatic equilibrium.

\appendix

\section{Multi-scalar field formulation}
\label{multi}

It is widely known that $f(R)$ theories can be rewritten as the Einsteinean gravity plus a scalar field model, with the curvature being dynamically identified with a scalar $\phi = R $ (or a function thereof) \cite{analogy}. Similarly, the discussed non-minimally coupled model \eq{model} is analogous to a multi-scalar field theory, with two scalar fields, $\phi = R$ and $\psi = {\cal L}$, as is illustrated below. One may perform a conformal transformation to the Einstein frame, so that the curvature term appears decoupled from these fields, and redefine the latter, obtaining the equivalent action

\beqa && S = \int  \sqrt{-g} d^4x \times \\ \nonumber && \bigg( 2 \ka \left[ R - 2g^\mn \si_{ij} \varphi^i_{,\mu} \varphi^j_{,\nu} -  4 
U(\varphi^1,\varphi^2) \right] + f_2(\varphi^2){\cal L^*} \bigg)~~, \eeqa

\noindent where $\varphi^1$ and $\varphi^2$ are scalar fields, $\si_{ij}$ is the field-metric

\beq \si_{ij} = \left(\begin{array}{cc}1 & 1 \\ -1 & 0\end{array}\right) ~~,\eeq

\noindent the potential is given by 

\beqa && U(\varphi^1,\varphi^2) = \\ \nonumber &&  {1 \over 4} \exp \left( -{2 \sqrt{3}\over 3} \varphi^1 \right) \left[\varphi^2 - 
{f_1(\varphi^2 )\over 2\ka }  \exp \left( -{2 \sqrt{3}\over 3} \varphi^1 \right)  \right]~~, \eeqa

\noindent and  ${\cal L^* } = \exp [-(4\sqrt{3}/3)\varphi^1]$.

The two scalar fields are related with the scalar curvature and the non-trivial $f_1(R)$ and $f_2(R)$ functions as

\beq \varphi^1 = {\sqrt{3}\over2} \log \left[ {F_1(R) + F_2(R) {\cal L} \over 2\ka} \right] ~~,\qquad  \varphi^2 = R ~~. \eeq

\noindent Inserting the expressions $f_1(R) = 2\ka \left(R + R^2/(6M^2)\right)$ and $f_2(R) = 1 + 2\xi R/M^2$, one gets

\beq  \varphi^1 = {\sqrt{3}\over2} \log \left[ 1 + {R + 6 \xi {\cal L} \over 3M^2} \right] ~~, \eeq

\noindent and the potential

\beqa && U(\varphi^1,\varphi^2) =   {1 \over 4} \varphi^2 \exp \left( -{2 \sqrt{3}\over 3} \varphi^1 \right) \times \\ \nonumber && \left[1 - 
\left(1 + {\varphi^2 \over 6M^2} \right)  \exp \left( -{2 \sqrt{3}\over 3} \varphi^1 \right)  \right]~~. \eeqa

\section{Evolution with trivial $f_1(R)$ term}

In this section, one attempts to obtain an inflationary solution to the modified Friedmann \eq{modFriedmann} by resorting only to the non-minimal coupling $f_2(R)$, {\it i.e.} setting $f_1(R) = 2\ka R$, its GR form, and taking a general non-minimal coupling $f_2(R)$. Considering only the dominant matter contribution of radiation, $\rho \approx \rho_r$, for simplicity, the Friedmann equation then becomes

\beqa \label{modFriedmann3} && \left( 1- F_2 {\rho \over 3\ka} \right)H^2 = \\ \nonumber && \left(f_2 + {F_2 R  \over 3}\right) {\rho \over 6\ka} + \left( {4 \over 3}{F_2^2 \over f_2}  - F_2' \right){\rho \over 3\ka} H\dot{R} ~~.  \eeqa

\noindent If one assumes that the effect of the non-minimal coupling dominates the usual GR terms, the above can be approximated by

\beq \label{Friedmannf2} - F_2 H^2 = {1 \over 2} \left(f_2 + {F_2 R  \over 3}\right) + \left( {4 \over 3}{F_2^2 \over f_2}  - F_2' \right) H\dot{R}~~.  \eeq

\noindent Remarkably, in this strongly coupled regime the density does not appear in the dynamical description of the cosmological evolution, and the obtained Friedmann Eq. is strikingly similar to the analogous expression arising from usual $f(R)$ theories \cite{review}. This observation paves the way to the possible use of the non-minimal coupling $f_2$ as the driving force behind inflation --- instead of  serving only to implement a reheating mechanism, with the inflationary regime occurring due to the $f_1(R)$ contribution.

\subsection{Power-law non-minimal coupling}

With the above discussion in mind, one first considers couplings of the form 

\beqa f_2(R) &=& 1 + \left({R \over M_2^2}\right)^n \gg 1 \rightarrow \\ \nonumber F_2(R) &\approx & {n \over R} f_2(R)~~, \eeqa

\noindent  assuming that the characteristic mass scale $M_2$ is well below the initial value for the scalar curvature, $R\gg M_2$. Clearly, the exponent $n$ must be positive, so that a decreasing curvature will eventually make the non-minimal coupling terms subdominant with respect to the GR contributions.

Substituting into \eq{Friedmannf2}, one gets

\beq H^2 = - \left( 1+ {n \over 3} \right) \left({R \over 2n} + {H \dot{R} \over R}\right)~~. \eeq

\noindent Notice that the mass scale $M_2$ vanishes from the above; resorting to the slow-roll approximation $R= 6H^2(2-\ep) \rightarrow \dot{R} \simeq -24H^3 \ep$, one may promptly obtain the slow-roll parameter

\beq \ep_n = { 36 + 18 n \over 36 + 27n + 4n^2} ~~. \label{epn} \eeq

\noindent The linear coupling $n=1$ yields $\ep_1 = 0.80 $, which (although smaller than unity) does not fulfill the slow-roll condition $\ep_n \ll1 $. Indeed, only for the unnatural range $n \gtrsim 40$ does one get $\ep_n \lesssim 0.1$. Also, notice that in the large $n$ regime one obtains $\ep_{\infty} = 9/2n $.

One now checks that the evolution of the radiation density $\rho$; using \eq{cons2} and the slow-roll approximation, one has

\beqa &&  \dot{\rho} = -\left(4H + {4 \over 3 } {F_2 \over f_2} \dot{R} \right) \rho =  \\ \nonumber && -\left(4H + {4n \over 3} {\dot{R} \over R} \right) \rho \simeq -4 \left(1 - {2n \over 3} \ep_n \right) H \rho = \\ \nonumber && -4 \left({36 + 3n - 8n^2 \over 36 + 27n +4n^2}\right) H\rho ~~, \eeqa

\noindent 
having used \eq{epn}. Only for $n < 2.32$ does one gets a diluting Universe, $\dot{\rho} < 0$; it is trivial to check that, for large $n$, $\dot{\rho} = 8 H \rho \rightarrow \rho \propto a^8$. Hence, the slow-roll range $n \gtrsim 40$ yields an exploding density, which would push back the expanding Universe into Planck scale concentration, rendering the model inapplicable.

Even if the density does not become radically high, another problem would arise. Firstly, it can be shown that, even though the density increases, the $f_2(R)$ derived terms in the modified Friedmann \eq{modFriedmann3} diminish: this occurs since the latter are of the generic form $F_2 R \rho$, and both $F_2$ and $R$ decrease. If this was not the case, then the Universe would become eternally dominated by the non-minimal coupling, even after the slow-roll condition $\ep \ll 1$ fails --- possibly including the so-called super-inflation, $\dot{H} > 0$.

However, this would only allow for inflation to end and the standard GR evolution to ensue: but this exit would be anything but graceful, as the Universe would be extremely hot, instead of at essentially zero temperature, overshooting the reheating temperature by several orders of magnitude. Hence, one concludes that no power-law models can solely drive a physically meaningful inflationary regime.

\subsection{Exponential non-minimal coupling}

Following the discussion after \eq{Friedmannf2}, one now resorts to an exponential non-minimal coupling function,

\beqa f_2(R) &=& \exp \left[ \left({R \over M_2^2}\right)^n \right] \gg 1 \rightarrow \\ \nonumber F_2(R) &\approx& {n \over R} \left({R \over M_2^2}\right)^n  f_2    ~~,\eeqa

\noindent again considering the coupled regime $R \gg M_2^2$. This is a somewhat natural choice, since it collapses to the trivial GR scenario for small values of ${\cal L}$, $f_2(R) \rightarrow 1$; as before, one considers a positive exponent $n$. Substituting into \eq{Friedmannf2}, one gets

\beqa H^2 = -{R \over 2} \left[ {1\over 3} + {1 \over n} \left({M_2^2 \over R}\right)^{n} \right] - \\ \nonumber \left[ {n \over 3} \left({R \over M_2^2}\right)^n +1 - n \right]  {H \dot{R} \over R} ~~. \eeqa 

\noindent Considering that $R \gg M_2^2$, the above simplifies to
 
\beq H^2 = -{R \over 6} - {n \over 3} \left({R \over M_2^2}\right)^n  {H \dot{R} \over R} ~~. \eeq

\noindent One uses the expansion

\beqa \left({R \over M_2^2}\right)^n &=& \left({6H^2(2-\ep) \over M_2^2}\right)^n \\ \nonumber &\simeq &  \left({12H^2 \over M_2^2}\right)^n \left(1 - {n\ep \over 2}\right) ~~, \eeqa

\noindent which, together with the usual slow-roll approximation, yields the slow-roll parameter

\beq \ep_n = {9 \over 2n} \left({M_2 \over 2\sqrt{3}H }\right)^{2n}  ~~, \label{epn2} \eeq

\noindent having again considered that $ H \gg M_2$. Given a sufficiently high initial Hubble parameter $H_i$, the the slow-roll condition $\ep_n \ll 1$ is clearly valid for a positive exponent $n$.

Inspection of \eq{epn2} shows that the slow-roll phase ends when

\beq H = f_n M_2 ~~~~, ~~~~f_n = 2\sqrt{3} \left({9 \over 2n}\right)^{1/2n} ~~, \eeq

\noindent with the numerical factor satisfying $f_n = 2\sqrt{3}$, for $n\rightarrow \infty$.

A striking result is that, if one takes $n=1$, {\it i.e.} a simple exponential coupling, then 

\beq \ep_1 = {3 \over 8} \left(M_2 \over H\right)^2~~,\eeq

\noindent which, from \eq{epsilon}, is identical to Starobinsky inflation, redefining $M_2 = 2M/3$.

As in the previous paragraph, one now ascertains what is the evolution of the radiation density $\rho$, following \eq{cons2}. As it turns out, the obtained result does not depend on the exponent $n$, as seen below:

\beqa && \dot{\rho} = - \left(4 H + {4 \over 3}{F_2 \over f_2} \dot{R} \right) \rho = \\ \nonumber && - 4 \left[H + {n \over 3} \left({R \over M_2^2}\right)^n {\dot{R} \over R} \right] \rho \simeq \\ \nonumber && - 4 \left[1 - {2n \over 3} \left({2\sqrt{3}H \over M_2}\right)^{2n} \ep_n \right] H\rho = 8 H\rho > 0 ~~.  \eeqa

\noindent Again, one encounters the same pathology already discussed: the density increases to pre-inflationary values; even if inflation ends before it reenters the Planck scale, one would be left with a burning hot Universe, much higher than the required reheating temperature.

As a final remark, one concludes that a suitable inflationary regime leading to a final reheating temperature appears unattainable resorting solely to a non-minimal coupling, for a wide choice of ``natural'' forms $f_2(R)$ (not shown in this manuscript), unless more model parameters are considered --- which would have to be introduced in an {\it ad-hoc} and unphysical fashion, in order to alleviate the above issue and tame the temperature evolution of the Universe.

Nevertheless, the mimicking behaviour exhibited in \eq{modFriedmann3}, attested by the recovery of Starobinsky inflation in the $f_2(R) = \exp (R/M_2^2)$ case, could perhaps be useful in the context of the early Universe, although none is envisaged in this study; a similar mimicking behaviour was obtained when considering the current accelerated expansion phase \cite{cosmo}.

The attempts described in this Appendix, although unsuccessful, serve to better frame the main purpose of this study: to show that a generalized non-minimal coupling may be paired to the Starobinsky quadratic curvature term in a modified Einstein-Hilbert action, so that the former successfully enables the preheating of the Universe, while it abstains from influencing the inflationary dynamics --- which is driven solely by the latter.

\end{document}